\documentclass[%
 reprint,
 amsmath,amssymb,
aps,
]{revtex4-1}

\usepackage{graphicx}
\usepackage{dcolumn}
\usepackage{bm}
\usepackage{subfigure}
\usepackage{color}
\usepackage{txfonts}
\usepackage{stmaryrd}
\usepackage{booktabs}
\usepackage{threeparttable}
\usepackage{multirow}
\usepackage{float}

\usepackage{soul}

\usepackage{geometry}
\usepackage{fancyhdr}
\begin{document}
\preprint{APS/123-QED}

\title{\textbf{Magnetism in doped infinite-layer NdNiO$_2$ studied by combined density functional theory and dynamical mean-field theory}}

\author{Dachuan Chen$^{1,2}$}
\email{These authors contributed equally to this work.}
\author{Peiheng Jiang$^{1}$}
\email{These authors contributed equally to this work.}
\author{Liang Si$^{3,4,1}$}
\author{Yi Lu$^{5,6}$}
\email{Corresponding author: yilu@nju.edu.cn}
\author{Zhicheng Zhong$^{1,7}$}
\email{Corresponding author: zhong@nimte.ac.cn}

\affiliation{
	$^1$CAS Key Laboratory of Magnetic Materials and Devices $\&$ Zhejiang Province Key Laboratory of Magnetic Materials and Application \\
     Technology, Ningbo Institute of Materials Technology and Engineering, Chinese Academy of Sciences, Ningbo 315201, China\\
	$^2$College of Materials Science and Opto-Electronic Technology, \\
	University of Chinese Academy of Sciences, Beijing 100049, China\\
    $^3$School of Physics, Northwest University, Xi'an 710069, China\\
	$^4$Institute for Solid State Physics, Vienna University of Technology, 1040 Vienna, Austria\\
	$^5$National Laboratory of Solid State Microstructures and Department of Physics, Nanjing University, Nanjing 210093, China\\
	$^6$Collaborative Innovation Center of Advanced Microstructures, Nanjing University, Nanjing 210093, China\\
	$^7$China Center of Materials Science and Optoelectronics Engineering, \\
	University of Chinese Academy of Sciences, Beijing 100049, China\
}

\date{\today}

\begin{abstract}
The recent observation of superconductivity in infinite-layer nickelates has brought intense debate on the established knowledge of unconventional superconductivity based on the cuprates. Despite many similarities, the nickelates differ from the cuprates in many characteristics, the most notable one among which is the magnetism. Instead of a canonical antiferromagnetic Mott insulator as the undoped cuprates, from which the superconductivity is generally believed to arise upon doping, the undoped nickelates show no sign of magnetic ordering in experiments. Through a combined density functional theory, dynamical mean-field theory, and model study, we show that although the increased energy splitting between O-$p$ orbital and Cu/Ni-$d$ orbital ($\Delta_{dp}$) results in larger magnetic moment in nickelates, it also leads to stronger antiferromagnetism/ferromagnetism competition, and weaker magnetic exchange coupling. Meanwhile, the self-doping effect caused by Nd-$d$ orbital screens the magnetic moment of Ni. The Janus-faced effect of $\Delta_{dp}$ and self-doping effect together give a systematic understanding of magnetic behavior in nickelates and explain recent experimental observations.
\end{abstract}

\maketitle
\thispagestyle{plain}
\cfoot{\thepage}

\section{\label{sec:level1}Introduction}

Since the discovery of high temperature superconductivity (HTSC) in cuprates \cite{bednorz1986possible}, enormous efforts have been devoted to understanding its microscopic mechanism and searching for new materials hosting HTSC \cite{tsuei2000pairing,scalapino2012common,lee2006doping}. The quasi-two-dimensional $\mathrm{CuO_2}$ planes in cuprates are generally regarded as a key element in realizing HTSC, whose antiferromagnetic (AFM) Mott insulating phase quickly melts away upon hole doping, following which the HTSC emerges \cite{lee2006doping,keimer2015quantum}. As the cuprates fall into the charge-transfer insulator category \cite{zaanen1985band}, the doped holes primarily reside on the O sites, which resonate around a Cu site, and their spins combine with the spin on Cu into a singlet \cite{zhang1988effective}. These Zhang-Rice singlets move around the lattice and form an effective one-band model, which serves as the starting point of many theoretical studies of HTSC \cite{lee2006doping}. With this general picture in mind, it is then natural to ask, if the isoelectronic infinite-layer nickelates $A$NiO$\mathrm{_2}$ with similar structural building blocks are also superconducting \cite{hayward1999sodium}?
	
This question has been answered recently by the discovery of superconductivity in hole-doped $\mathrm{Nd_{0.85}Sr_{0.15}NiO_{2}}$ with $T_c\sim$15 K by Li $\emph{et. al.}$ \cite{li2019superconductivity,li2020superconducting} and confirmed later by several other groups \cite{zeng2020phase,gu2020single,zhou2021antiferromagnetism}. Despite the great similarities between the nickelates and cuprates, stark differences remain, especially in their magnetic properties, which raises questions regarding the validity of a uniformed description of their superconductivity. Neutron diffraction studies show that the magnetic moment of cuprates is about 0.5 $\mu_B$  \cite{vaknin1989antiferromagnetism}. However, based on the hypothetical G-type AFM structure, neutron diffraction experiments indicate that the magnetic moment of nickelates (0.05 $\mu_B$/Ni \cite{hayward1999sodium} or 0.06 $\mu_B$/Ni \cite{hayward2003synthesis}) is much smaller than that of cuprates. While several first-principle studies propose that the AFM order is the ground state in nickelates \cite{anisimov1999electronic,lee2004infinite,been2021electronic,leonov2020lifshitz,botana2020similarities,choi2020role,kapeghian2020electronic,lechermann2020late,si2020topotactic,liu2020electronic,zhang2020effective,ryee2020induced}, experiments have so far failed to obtain direct evidence of long-range magnetic order in them \cite{hayward1999sodium,hayward2003synthesis}.
	
On the other hand, short-range AFM interactions are observed in NMR \cite{yi2021nmr}, Raman spectroscopy \cite{fu2019core}, and Resonant Inelastic X-Ray Scattering (RIXS) \cite{lin2021strong,lu2021magnetic} experiments. Such a magnetic behavior is also reported by plenty of theoretical studies \cite{liu2020electronic,zhang2020effective,katukuri2020electronic,ryee2020induced,nomura2020magnetic}, some of which further suggest that the exchange interaction of nickelates is not stronger than that of cuprates \cite{liu2020electronic,zhang2020effective}.

In this paper, we focus on the origin of the magnetic differences between the cuprates and nickelates, and perform a systematic study on their magnetic properties with both density functional theory plus Hubbard $U$ (DFT+$U$) and dynamical mean-field theory (DMFT). Our calculations show several important consequences of the large $\Delta_{dp}$ and self-doping effect in nickelates. First, the energy difference between various magnetic orders is quite small in undoped nickelates, leading to the competition of different magnetic orders and even possible spin fluctuation \cite{ortiz2021magnetic,huangfu2020short}. Second, while the AFM state is energetically more favorable in undoped and electron doped nickelates, an intralayer ferromagnetic (FM) state is surprisingly stabilized upon hole doping. Third, we found that the magnetic moment of Ni is larger than that of Cu. The last observation is reconciled with the seemingly contradicting experimental results by futher considering the self-doping and dynamic screening effects in nickelates using DMFT. In addition, our exact diagonalization (ED) calculation of small clusters reports a smaller magnetic exchange interaction $J$ in nickelates, which is still comparable to that in the cuprates.

\section{\label{sec:level2}Methods}
	
	\emph{DFT.} We performed DFT calculations within generalized gradient approximation Perdew-Burke-Ernzerhof (GGA-PBE) functional and the projector augmented wave (PAW) method as implemented in the Vienna \emph{ab initio} simulation package (\textsc{vasp}) \cite{kresse1996,perdew1996generalized,kresse1996efficient} and \textsc{Wien2K} \cite{blaha2001wien2k}. In addition to $\mathrm{NdNiO_2}$, the isostructural $\mathrm{CaCuO_2}$ is also studied for comparison. The Nd-$4f$ orbitals are treated as core states. The relaxed lattice constants are $a = b$ = 3.88 {\AA} and $c$ = 3.35 {\AA} for $\mathrm{NdNiO_2}$, and $a = b$ = 3.81 {\AA} and $c$ = 3.17 {\AA} for $\mathrm{CaCuO_2}$. The strain effect, which is discussed in Appendix A, does not affect the conclusions of this work. Electronic correlations on Ni/Cu-$3d$ and Nd-$5d$ are included by DFT+$U$ method with $U$ = 5.0 eV and 2.0 eV, respectively. We also adopted Heyd-Scuseria-Ernzerhof (HSE06) hybrid functional for calculations, because this functional performs exceptionally well in describing systems with a mixture of itinerant and localized electrons from different orbitals \cite{krukau2006influence}. The strongly constrained and appropriately normed (SCAN) functional which enables improved predictions for oxides \cite{sun2015strongly} is also included in calculations. To determine the magnetic ground state, various possible magnetic orders are considered. Since the interlayer interactions are weaker than the intralayer ones, here we mainly focus on intralayer magnetic order, including FM (0,0,0), C-type AFM ($\pi,\pi,0$) ($\sqrt{2}\times\sqrt{2}\times1$ supercell) and stripy AFM ($\pi,0,0$) (S-type, $2\times2\times1$ supercell). Some other complex orders are also carefully checked (see Appendix B). The kinetic energy cutoff is set to 500 eV. The Brillouin zone is sampled with $13\times13\times15$, $8\times8\times12$ and $8\times8\times8$ Monkhorst-Pack $\mathbf{k}$-grids for PBE calculations of unit cell and supercell, and HSE06 calculations, respectively. The doping is simulated by using virtual crystal approximation (VCA). We also calculated Sr-doping by constructing supercell of NdNiO$_2$, which shows similar results with VCA scheme.
	
	\emph{Exact Diagonalization.} To obtain the exchange interaction $J$, we construct a three-band Hubbard model in a $\mathrm{Cu_5O_{16}}$ cluster relative to a Cu-$d^{10}$, O-$p^6$ vacuum state, and map the low-energy spectrum to that of a Heisenberg model. For simplicity, the Coulomb interaction paramters are kept the same for both the nickelate and cuprate with values listed in Ref.\cite{hybertsen1990renormalization}. The calculation is performed using \textsc{Quanty} \cite{haverkort2012multiplet}.	
	
	\emph{DMFT.} For the DMFT calculations, the DFT band structure around the Fermi level is projected onto Wannier functions \cite{wannier1937structure,marzari2012maximally} using \textsc{Wien2Wannier} \cite{mostofi2008wannier90,kunevs2010wien2wannier} and supplemented by a local density-density interaction. To study the magnetic properties of $\mathrm{NdNiO_2}$ with and without self-doping effect, we choose Ni-$d$ orbitals as the correlated space to calculate the magnetic phase diagram and conductivity. The self-doping effect is then included by modifying electron fillings on the DMFT level. To study the local screening effect from Nd, we also construct a Nd-$d$+Ni-$d$ model with ten bands. The interaction parameters are computed by constrained random phase approximation (cRPA) \cite{miyake2008screened,sakakibara2020model,nomura2019formation}, which gives an averaged inter-orbital interaction $U^\prime$ = 3.10 eV and Hund's exchange $J$ = 0.65 eV for Ni. The intra-orbital Hubbard interaction follows as $U=U^\prime + 2J$. The resulting Hamiltonian is then solved in DMFT using continuous-time quantum Monte Carlo simulations in the hybridization expansions \cite{gull2011continuous} implemented in w2dynamic \cite{parragh2012conserved,wallerberger2019w2dynamics}. The maximum entropy method \cite{gubernatis1991quantum,sandvik1998stochastic} is employed for analytic continuation of the spectra.

\section{\label{sec:level3}Results and Discussions}	
\subsection{\label{sec:level4}DFT calculations}

\begin{figure}[htbp]
	\centering
	\includegraphics[width=0.45\textwidth]{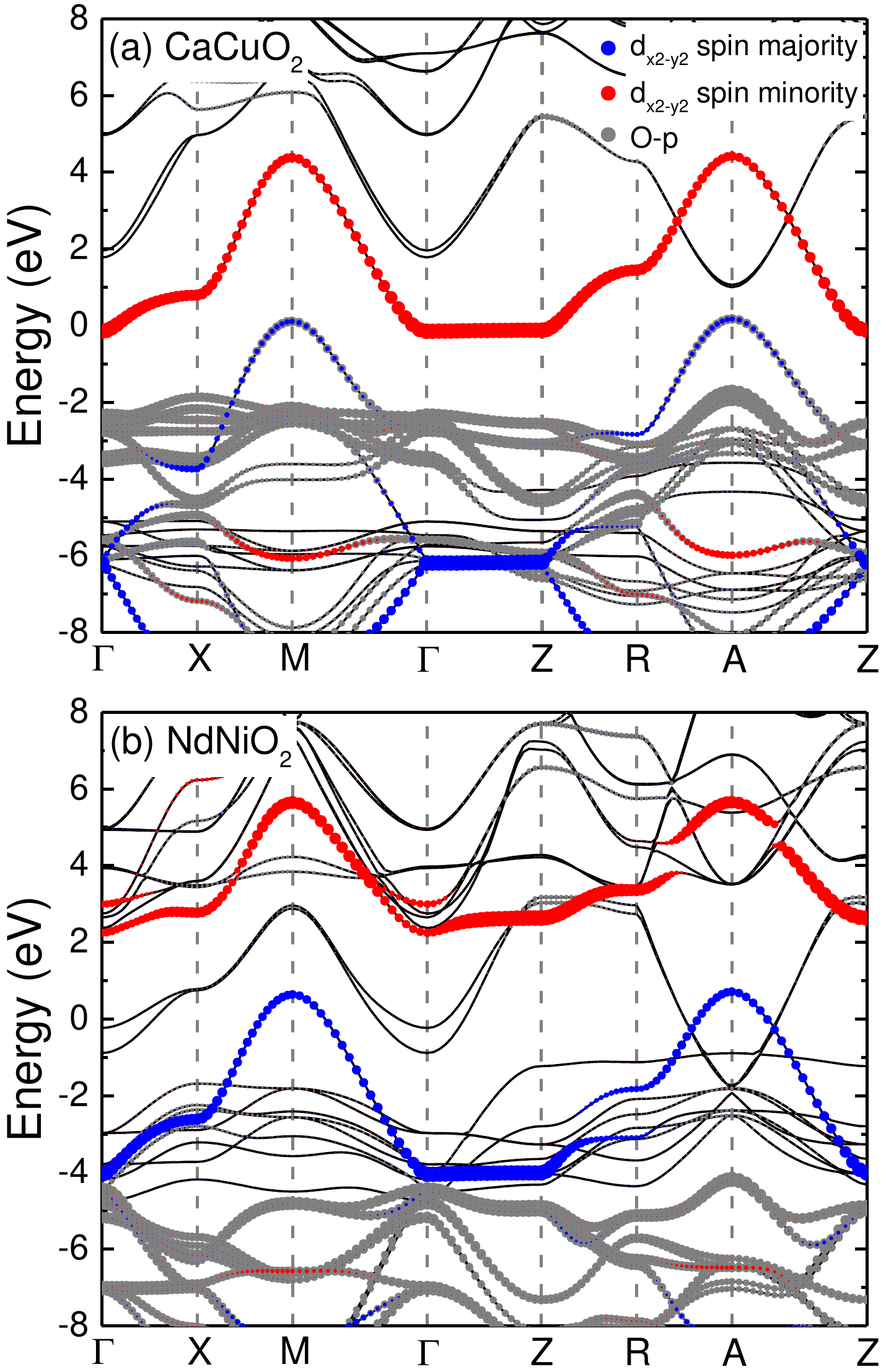}
	\caption{Electronic band structures of (a) $\mathrm{CaCuO_2}$ and (b) $\mathrm{NdNiO_2}$ with FM order calculated by HSE hybrid functional.}
	\label{fig1}
\end{figure}

\begin{figure*}[htbp]
	\centering
	\includegraphics[width=0.8\textwidth]{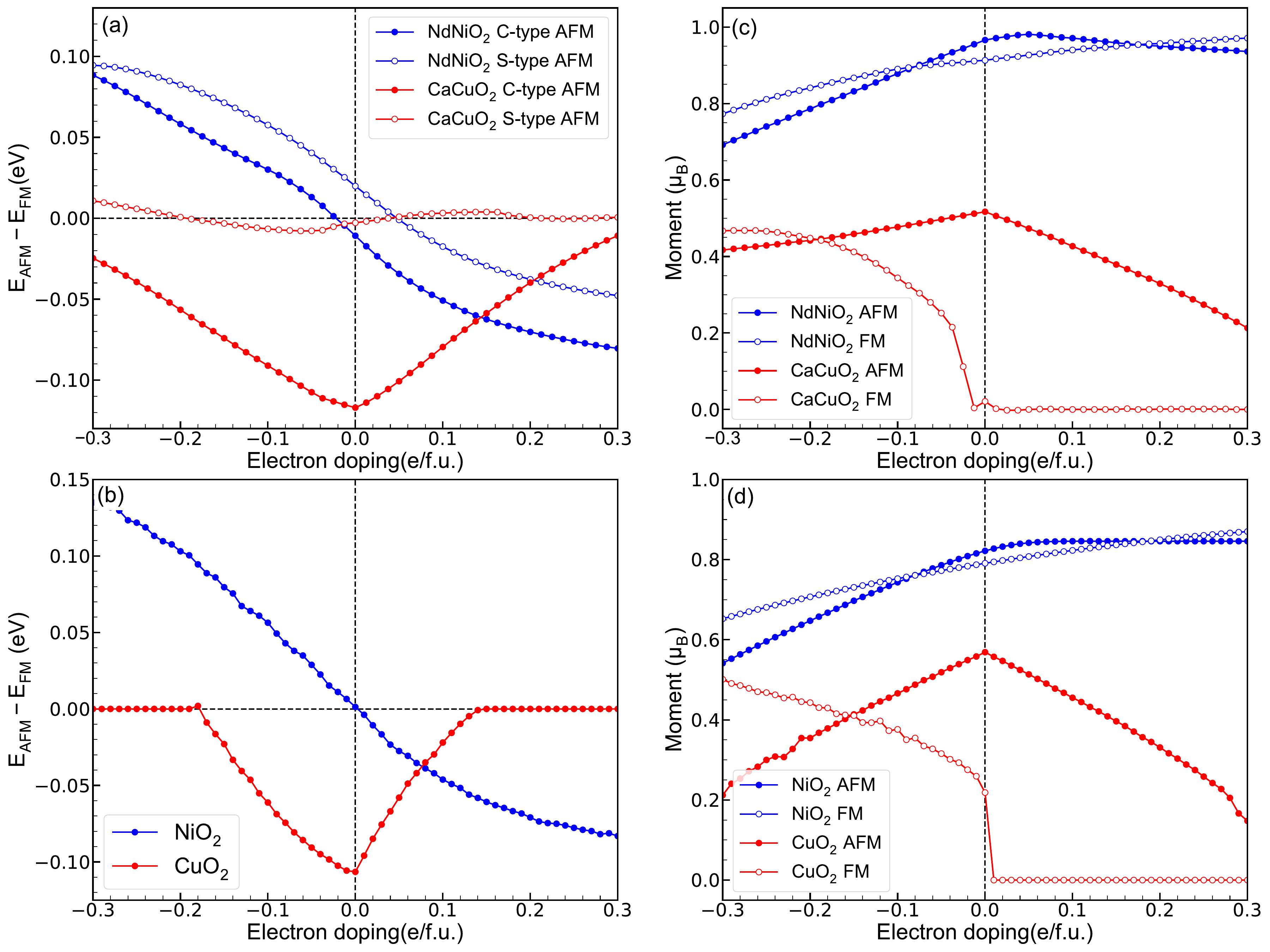}
	\caption{(a-b) Total energy difference between AFM and FM magnetic orders at different doping concentrations with (a) DFT calculations and (b) mean-field solutions. The electron and hole doping are represented by positive and negative doped electrons, respectively. The energy of FM order is set to zero. (c-d) Magnetic moment of $\mathrm{NdNiO_2}$ and $\mathrm{CaCuO_2}$ at different doping concentrations with (c) DFT calculations and (d) mean-field solutions.}
	\label{fig2}
\end{figure*}

The HSE calculated electronic band structures of $\mathrm{CaCuO_2}$ and $\mathrm{NdNiO_2}$ with FM configuration are shown in Fig. 1. We first focus on the spin splitting of Cu/Ni-$d$ orbital and find that the splitting of $\mathrm{NdNiO_2}$ is noticeably larger than $\mathrm{CaCuO_2}$. Especially for the $d_{x^2-y^2}$ orbital near Fermi level, the spin splitting of $\mathrm{NdNiO_2}$ is 0.74 eV larger than $\mathrm{CaCuO_2}$ at $M$ point (5.01 eV for $\mathrm{NdNiO_2}$ and 4.27 eV for $\mathrm{CaCuO_2}$). This indicates a larger local moment of Ni than that of Cu. Indeed, the projected moment is 0.94 $\mu_B$ for Ni in $\mathrm{NdNiO_2}$ and 0.72 $\mu_B$ for Cu in $\mathrm{CaCuO_2}$. We also calculate the electronic band structure and moment of $\mathrm{CaCuO_2}$ and $\mathrm{NdNiO_2}$ with PBE+$U$ and SCAN functional, both of them give similar results. The larger magnetic moment of $\mathrm{NdNiO_2}$ can be explained by the suppressed moment screening, which is determined by the energy splitting between Ni/Cu-$d_{x^2-y^2}$ and O-$p$ orbitals ($\Delta_{dp}$). A large (small) $\Delta_{dp}$ indicates a weak (strong) hybridization between O-$p$ and Ni/Cu-$d_{x^2-y^2}$ orbitals, and hence weak(strong) screening and large (small) magnetic moment. The calculated $\Delta_{dp}$ of $\mathrm{NdNiO_2}$ is 2$\sim$3 eV larger than that of $\mathrm{CaCuO_2}$, whose exact value also depends on calculation method and detail \cite{jiang2019electronic,hansmann2014importance}. Based on the calculations, we find that the larger spin splitting and $\Delta_{dp}$ contribute to a larger local moment in nickelate than in cuprate, so that the pristine nickelate is not nonmagnetic (NM).

In addition to the large local moment, another key factor for generating long-range magnetic order is magnetic interaction. Without strong magnetic interaction, the system will converge to a NM state. To investigate the interaction, we further calculate various magnetic orders and their doping dependence in both the nickelate and cuprate. Fig. 2(a) shows the DFT+$U$ calculated energy difference between various magnetic orders upon electron/hole doping. For $\mathrm{CaCuO_2}$, the ground state is in a stable C-type AFM state for the undoped case. When electron and hole doping concentrations increase from 0 to 0.3 /f.u., the energy difference between C-type AFM and FM (the DFT+$U$ calculated FM order converges to NM order with $U_{Cu}$ = 5 eV when the doping concentration is larger than 0.01 e/f.u.) decreases from 120 meV to 25 meV for hole doping and 11 meV for electron doping. This result indicates that both electron and hole doping can suppress AFM coupling of $\mathrm{CaCuO_2}$, leading to the disappearance of long-range magnetic order. For the undoped $\mathrm{NdNiO_2}$, the energy difference between AFM and FM orders is $-$11 meV for C-type AFM and 25 meV for S-type AFM, such a small energy difference implies that a competition exists between these magnetic orders. With electron doping larger than 0.02 /f.u., the C-type AFM state has the lowest energy, which is about 30 meV or 80 meV lower than S-type AFM or FM state, respectively. In the case of hole doping, a striking intralayer FM state can be induced in $\mathrm{NdNiO_2}$, which becomes more stable when the hole doping concentration increases, the energy different between FM and AFM can be as large as 90 meV with 0.3 /f.u. hole doping. We note that signs of ferromagnetic order have also been observed experimentally in overdoped cuprates \cite{sarkar2020ferromagnetic,sonier2010direct}.

The DFT+$U$ calculated local magnetic moment of doped $\mathrm{NdNiO_2}$ and $\mathrm{CaCuO_2}$ are shown in Fig. 2(c). For CaCuO$_2$, the moment maximizes at half filling with a value of 0.5 $\mu_B$, and gradually decreases with increasing electron or hole doping.  For $\mathrm{NdNiO_2}$, due to the self-doping effect induced by Nd-$d$ orbitals, the maximal moment appears after slight electron doping. For both hole and electron doping, the moments of $\mathrm{NdNiO_2}$ are always larger than $\mathrm{CaCuO_2}$, which is consistent with the large spin splitting in nickelate, and is contributed by self-doping effect and large energy gap $\Delta_{dp}$ as discussed above. To check the reliability of these results, we also perform HSE calculations which give similar conclusions with DFT+$U$ scheme.

Our DFT calculated result is in agreement with the well-known fact that the undoped cuprate is a Mott insulator. Its ground state is intralayer AFM, and this long-range magnetic order is well established by Cu-O-Cu super-exchange interaction. Doping can suppress the AFM order and close the gap, and then drive pesudogap and finally superconducting phase. For nickelate, the large $\Delta_{dp}$ reduces Ni-O-Ni super-exchange interaction and weaken the AFM order. Moreover, the self-doping effect contributes a large kinetic part, and then results in a metallic state which favors spin parallel. Therefore, both of these two effects driving $\mathrm{NdNiO_2}$ to be in a competing magnetic phase.

\subsection{\label{sec:level5}Mean-field solution of Emery model}

\begin{figure*}[htbp]
	\centering
	\includegraphics[width=0.8\textwidth]{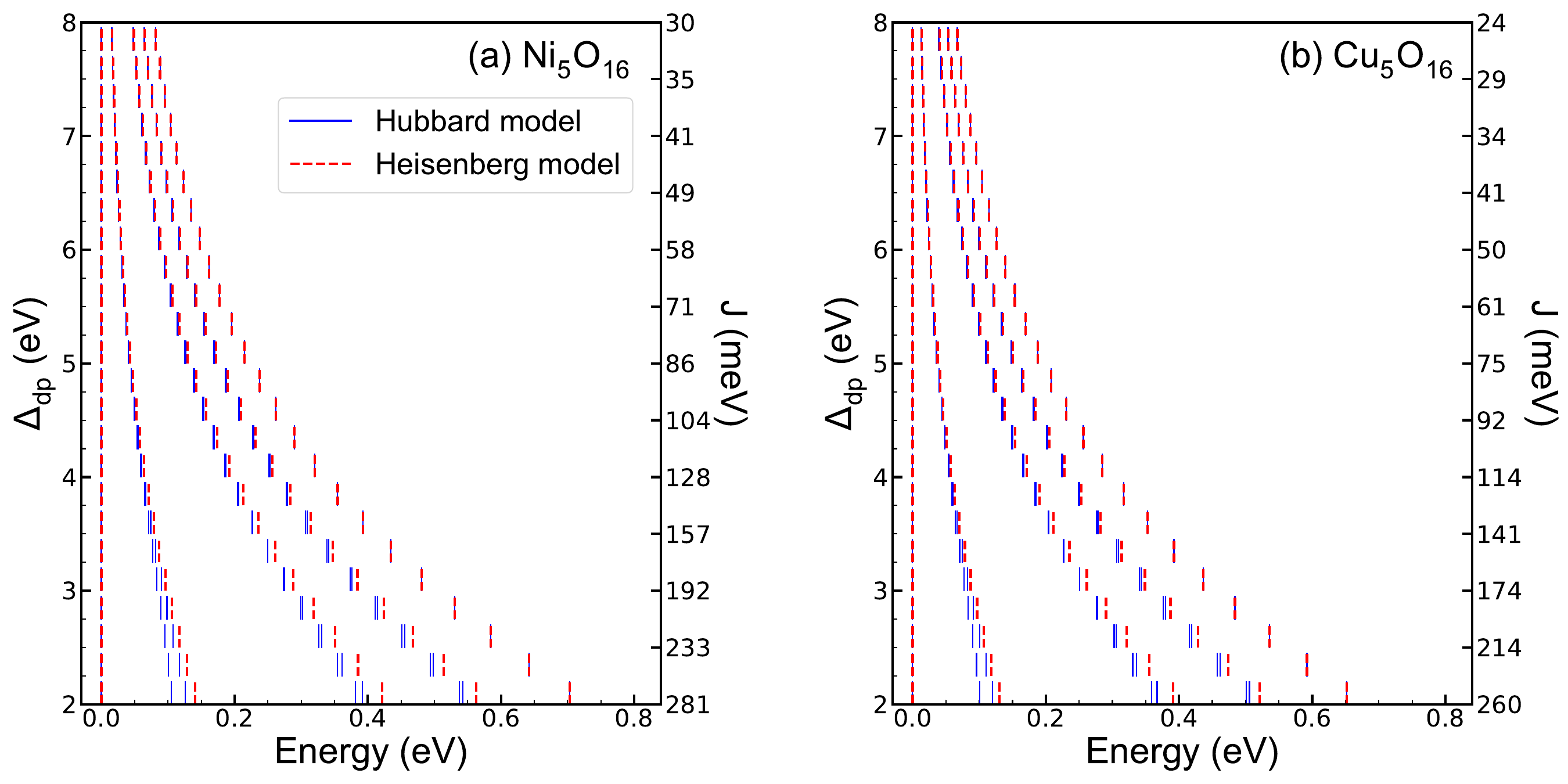}
	\caption{The low-energy spectrum for $\mathrm{Cu_5O_{16}}$ and $\mathrm{Ni_5O_{16}}$ clusters calculated in the three-band Hubbard model (blue line) in comparison to mappings onto effective Heisenberg model (red dash line) for half filled case.}
	\label{fig3}
\end{figure*}
	
To further confirm that the magnetic difference between nickelate and cuprate mainly comes from $\Delta_{dp}$ and doping effect, we construct a minimal three-orbital Hubbard model on the two dimensional $\mathrm{CuO_2}$/$\mathrm{NiO_2}$ plane, that captures the relevant physical degrees of freedom,

\begin{equation}
\begin{split}
	{\cal H}=&t_{dp}\sum_{<ij>\sigma}d^\dagger_{i\sigma}p_{j\sigma}+t_{pp}\sum_{<ij>\sigma}p^\dagger_{i\sigma}p_{j\sigma}+
	t_{dd}\sum_{<ij>\sigma}d^\dagger_{i\sigma}d_{j\sigma}\\
	&+\Delta_{dp}\sum_{i\sigma}p^\dagger_{i\sigma}p_{i\sigma}+U_d\sum_{i}n^\uparrow_{di}n^\downarrow_{di} ,
\end{split}
\end{equation}

\noindent where $d^\dagger_{i\sigma}$ and $p^\dagger_{i\sigma}$ are creating operators of spin $\sigma(\uparrow,\downarrow)$ at $i$-th site of $d_{x^2-y^2}$ and $p_{x/y}$ orbitals, respectively. For simplicity, we include Coulomb interaction only for $d_{x^2-y^2}$ orbital. The hopping parameters $t_{dp}$, $t_{pp}$, $t_{dd}$ and $\Delta_{dp}$ are taken from Ref. \cite{hansmann2014importance}. For cuprate, $t_{dp}$ = 1.48 eV, $t_{pp}$ = 0.91 eV, $t_{dd}$ = 0.15 eV, and $\Delta_{dp}$ = 0.95 eV, while for nickelate, we only change $\Delta_{dp}$ to 4.45 eV. Because the local interaction of Ni-$d$ orbital is similar with Cu-$d$, we use the Hubbard interaction $U_d^{Ni}=U_d^{Cu}=4$ eV (Note that to be consistent with DFT calculations and take into account the overestimation of polarization in mean-field solution, we set a relatively small Hubbard $U$). The Nd-$d$ band is excluded in the calculation to minimize the number of parameters in the model. To account for its self-doping effect, the zero-doping reference point is shifted by 0.15 hole/f.u. according to DFT calculation \cite{deng2021first}.

For comparison, we calculate the total energy and magnetic moment of $\mathrm{CuO_2}$/$\mathrm{NiO_2}$ planes with different electron concentrations within the Hatree Fock mean-field approximation. Fig. 2(b) shows the energy difference of AFM and FM state ($E$$\mathrm{_{AFM}}$$-$$E$$\mathrm{_{FM}}$). The value of energy difference is set to zero when the ground state is paramagnetic (PM). For cuprate, since the $\Delta_{dp}$ in cuprate is small, the adjacent spins on Cu atom show strong AFM coupling through the super-exchange interaction, which is proportionate to $1/\Delta_{dp}^2$. The relatively large superexchange coupling leads to AFM order when undoped. As for nickelate, the AFM coupling is weakened due to the larger $\Delta_{dp}$. The kinetic energy of the extra holes in the undoped nickelate further frustrate the AFM order. Based on this mechanism, nickelate and cuprate exhibit completely different behavior in their ground states. Cuprate shows a stable AFM in parent compound, and it is suppressed upon doping. While nickelate shows a FM and AFM competition phase when the hole concentration increases. These complicated magnetic competition phases, such as spin spirals, nematicity and FM domain walls, have been discussed on mean-field level \cite{chiciak2018magnetic}. Especially with hole doping, both DFT calculations and model analysis give the same intralayer FM ground state configuration. Please note that both DFT and model analysis are on the mean-field level, which always overestimate magnetic order. For the effect of Nd-$d$ orbitals, the self-doping effect is well known, actually, some group suggest the hybridization effect is also important \cite{gu2020substantial}, which may induce Kondo context \cite{yang2022self} and/or may contribute to RKKY coupling between Ni moments as suggested by recent experiment \cite{lu2021magnetic}.

Fig. 2(d) shows the calculated magnetic moment. The key difference between cuprate and nickelate is the larger $\Delta_{dp}$ and the self-doping effect in nickelate. The large $\Delta_{dp}$ in nickelate induces strong spin polarization in $d$ orbital and thus a larger local moment. In cuprate, the strong $d$-$p$ hybridization resulted from small $\Delta_{dp}$ makes $d$ orbital difficult to be strongly polarized. In addition, due to the self-doping effect, the magnetic moments of cuprate and nickelate show completely different behavior under doping. In nickelates, the additional electrons occupy the Nd-$d$ orbitals, hence the moment tends to saturate gradually under electron doping. On the contrary, the moment in cuprate will decrease rapidly since there is no self-doping band and the O-$p$ orbital will also suppress spin polarization especially in FM state. The overall doping behavior of the energy difference between AFM and FM states as well as the moment are in great agreement with our DFT+$U$ results, and confirming our initial conclusions that the $\Delta_{dp}$ and self-doping as the key factors responsible for the magnetic difference between the nickelates and cuprates.

\subsection{\label{sec:level6}Exact Diagonalization}

\begin{figure*}[htbp]
	\centering
	\includegraphics[width=0.9\textwidth]{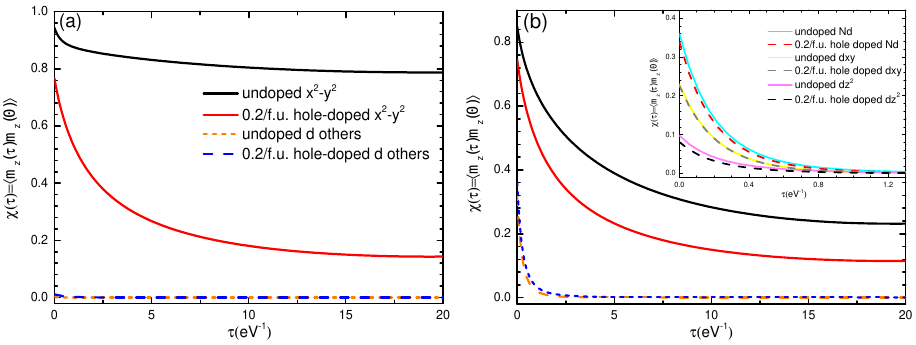}
	\caption{Orbital resolved local spin correlation functions $\chi(\tau)=\left\langle\hat{m_z}(\tau)\hat{m_z}(0)\right\rangle$ calculated by DMFT at $T$ = 300  K in the Kanamori form, $\beta/2 \approx 19.34$ \ eV$\mathrm{^{-1}}$. (a) Ni-$d$ only model, (b) Nd-$d$ + Ni-$d$ model, inset: the contributions of Nd atom. The black curve in (b) shows a reduction compared with that in (a) due to the self-doping effect. }
	\label{fig4}
\end{figure*}

In order to evaluate the exchange interaction's contribution to superconductivity, we further calculate the magnetic exchange interaction $J$ between the Ni local magnetic moment in nickelate, and compare it to cuprate. Based on the DFT and mean-field results above, if we simply map the total magnetic energy to Heisenberg model to calculate the exchange interaction ($J\propto$ $E$$\mathrm{_{AFM}}$$-$$E$$\mathrm{_{FM}}$), the calculated $J$ will be close to zero because of the competition between exchange interaction which favors AFM and kinetic part which favors FM. Such an energy mapping method is applicable in undoped cuprates since the undoped cuprates are insulator. As for nickelates, the metallic behavior emerges because of the itinerant holes are invovled by self-doping bands, and then the kinetic energy contributes significantly to a total energy. Thus, to obtain the exchange interaction in a metal, one must construct a equivlent substitution which can be treat as an insulator (see Ref. \cite{nomura2020magnetic}), or include the kinetic term in calculations as ED method. Now we follow the method in Ref. \cite{hybertsen1990renormalization,zhang2020effective}, and construct a three-band Hubbard model on $\mathrm{Cu_5O_{16}}$ and $\mathrm{Ni_5O_{16}}$ clusters. The parameters for $\mathrm{Ni_5O_{16}}$ are consistent with cuprate except a smaller $U_d^{Ni}$=8 eV. To introduce the boundary interacitons, we embed the clusters in an array of Cu/Ni-$d^9$ sites, which shifts the effective on-site energy of the outer O orbitals due to the intersite Coulomb energy. From the ED studies of half-filled $\mathrm{Cu_5O_{16}}$ and $\mathrm{Ni_5O_{16}}$ clusters, we obtain the low-energy spectra with different $\Delta_{dp}$ values. These spectra are matched with those of a nearest-neighbor spin-1/2 Heisenberg model ($H=\sum_{\langle ij\rangle }J S_iS_j$) with different exchange coupling $J$. As Fig. 3 shows, exchange coupling $J$ is very sensitive to $\Delta_{dp}$, but it is nearly not dependent on Coulomb interaction $U_d$. In the $\mathrm{Cu_5O_{16}}$ cluster, $J$ is about 135 meV when $\Delta_{dp}$=3.6 eV. While in $\mathrm{Ni_5O_{16}}$ cluster, $\Delta_{dp}$ is 1.9 eV to 2.7 eV larger than $\mathrm{Cu_5O_{16}}$, so that the exchange coupling $J$ in $\mathrm{Ni_5O_{16}}$ cluster is about 62$\pm$9 meV. Thus, the effective AFM spin-spin interaction between Ni local spin is not weak. The long-range magnetic order can not be stabilized because of the large kinetic term, which can be treated as $t$-$J$ model.

\subsection{\label{sec:level7}DMFT calculation}

Our DFT+$U$ calculations indicate a larger local magnetic moment of Ni in $\mathrm{NdNiO_2}$ than Cu in $\mathrm{CaCuO_2}$ and a strong magnetic order in $\mathrm{NdNiO_2}$. However, neutron diffraction experiments failed to observe sizable magnetic moments in the nickelates \cite{hayward1999sodium,hayward2003synthesis}. To explain this contradiction, we employ DFT+DMFT method to investigate the dynamical screening processes of magnetic moment in nickelates.

To study the local moment of $\mathrm{NdNiO_2}$ within DFT+DMFT, we compute the local spin correlation functions $\chi(\tau)=\left\langle\hat{m_z}(\tau)\hat{m_z}(0)\right\rangle$. We first consider a Ni-$d$ only model as shown in Fig. 4(a), $\chi(0)$ of $d_{x^2-y^2}$ orbital without doping is about 0.9 $\mu_B^2$. Upon 0.2 /f.u. hole doping, $\chi$(0) is slightly reduced to 0.76 $\mu_B^2$ and quickly decreases to 0.15 $\mu_B^2$ when $\tau = \beta/2$ ($\tau$ = 19.34 eV$\mathrm{^{-1}}$). The observed fast decay of $\chi$ at finite $\tau$ reflects a dynamical screening of the local moment due to spin fluctuations. The other orbitals, including $t$$_{2g}$ and $d$$_{z^2}$, have no contributions to magnetic moment.

To consider the Nd contribution, we then construct a Ni-$d$ + Nd-$d$ model with total 10 orbitals. As Fig. 4(b) shows, the instantaneous spin correlation functions of the nominally undoped Ni-$d_{x^2-y^2}$ orbital is reduced to 0.8 $\mu_B^2$ which is consistent with our previous DFT+$U$ results. This size of moment indicates the occupations in Ni-$d_{x^2-y^2}$ orbital is slightly less than half-filling. We see that when $\tau = \beta/2$, the spin correlation functions of undoped Ni-$d_{x^2-y^2}$ orbital is screened to 0.24 $\mu_B^2$, which is similar to the hole doped case in $d$-only model. Meanwhile, other Ni-$d$ orbitals have a considerable instantaneous spin correlation functions ($\mathrm{\thicksim}$ 0.36 $\mu_B^2$) and rapidly reduced to 0 for $\tau>$ 0. The moment is further screened by doping which is the same with Ni-$d$ only model. These phenomena indicate the introduction of Nd-$d$ orbitals play a charge reservoir role, resulting the decrease of occupations in Ni-$d_{x^2-y^2}$, leading to a smaller moment. When 0.2 /f.u. hole doping is applied, the spin correlation functions of Ni-$d_{x^2-y^2}$ orbital is reduced to about 0.12 $\mu_B^2$, and other orbitals exhibit the same behavior as the undoped case. The contribution of Nd-$d$ orbitals are shown in the inset of Fig. 4(b). The instantaneous spin correlation functions of undoped and 0.2 /f.u. hole doped Nd-$d$ orbitals are 0.36 $\mu_B^2$ and 0.34 $\mu_B^2$, respectively, and the spin correlation functions of all these orbitals falls rapidly to 0 when $\tau$ increases. The difference between Ni-$d$ only model and Nd-$d$ + Ni-$d$ model is that the Ni-$d_{x^2-y^2}$ moment has been screened by Nd-$d$ orbitals due to the self-doping effect without additional doping. DMFT benchmarks with density-density Hamiltonian gives the similar consequence as shown in Fig. 9 in Appendix D. Our DFT+DMFT calculations suggest that self-doping effect in nickelates induces screening effect to Ni local moment, this reduces the time-average magnetic moment. In this respect, the frequency range of typical neutron experiments would be too limited to directly measure the magnitude of the moment of $\mathrm{NdNiO_2}$.

\begin{figure*}[htbp]
	\centering
	\includegraphics[width=0.6\textwidth]{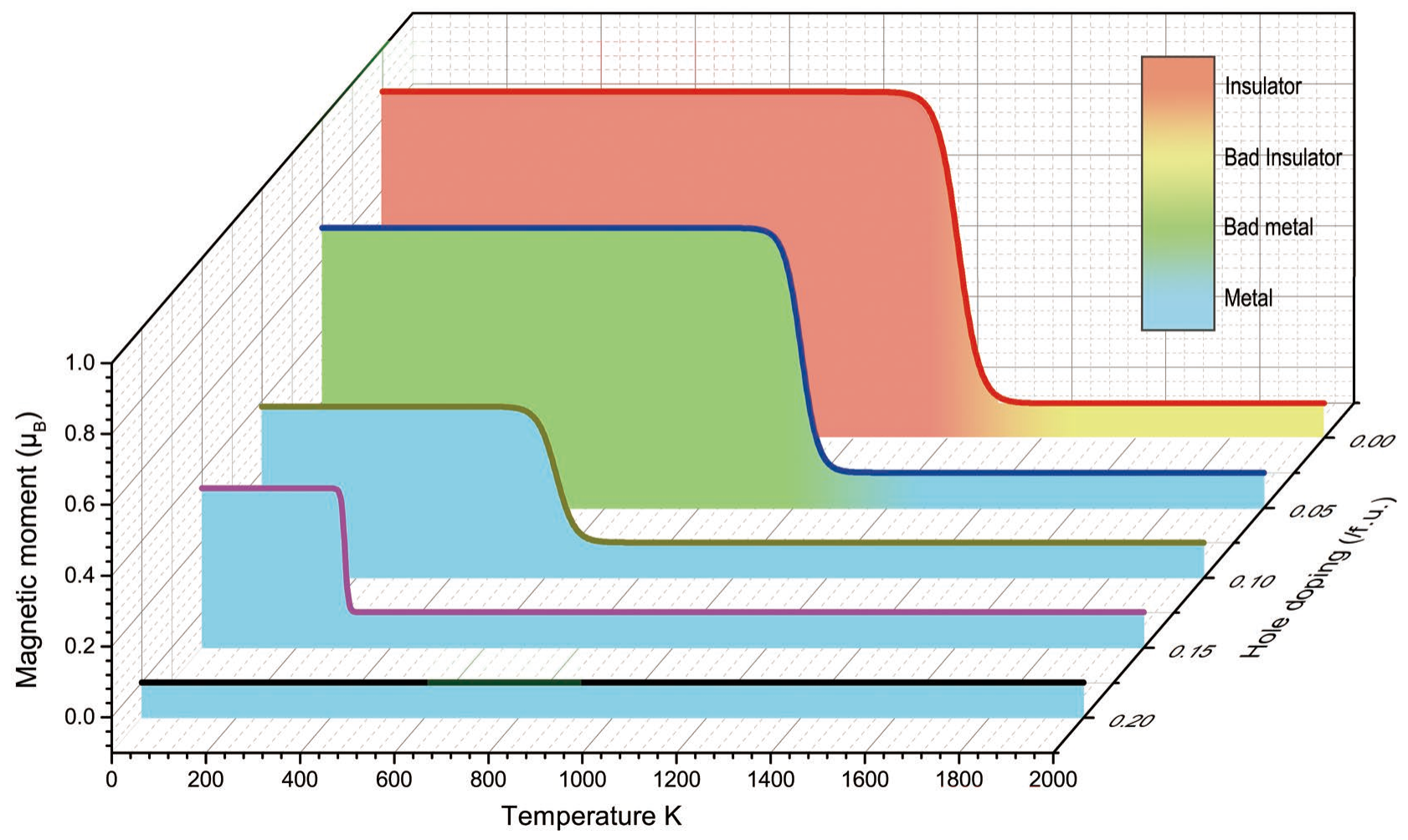}
	\caption{Phase diagram of $\mathrm{NdNiO_2}$ calculated by 2-site Ni-$d$-only DMFT method, which shows the magnetic moment and conductivity at different temperature and hole doping concentration. The black, gray, mazarine and blue circle represent AFM insulator, AFM metal, PM insulator and PM metal, respectively. The phases of bad insulator and bad metals are defined by \cite{rozenberg1995optical}. }
    \label{fig5}
\end{figure*}

We further investigate the temperature dependence of magnetic moment of $\mathrm{NdNiO_2}$ within hole doping range from 0 to 0.2 /f.u. by studying a two-site Ni-$d$ model. As Fig. 5 shows, without the inclusion of the Nd-$d$ band, undoped $\mathrm{NdNiO_2}$ has an AFM insulating ground state with high N\'{e}el temperature, in stark contrast with experimental findings. This highlights the importance of Nd-$d$ band and its self-doping effect in determining the low-energy properties of the nickelates. We note that in addition to self-doping, several groups also emphasize on the importance of hybridization between the Nd-$d$ and Ni-$d$ \cite{zhang2020self,gu2020substantial,yang2022self}. Based on this and other earlier observations made in this work and the hybridization analysis in previous works \cite{jiang2019electronic,kitatani2020nickelate,karp2020many}, we thus emphasize that an appropriate minimal model of the nickelate should include the self-doping effect of Nd-$d$, thus clearly distinguishing the nickelate from the cuprate. In our work, the local moment is about 0.95 $\mu_B$ at low temperature, which decreases to zero when the temperature is above N\'{e}el temperature. The N\'{e}el temperature is calculated to be about 1200 K, which is significantly higher than the experimental value due to the well-known DMFT overestimation of the magnetic transition temperature of AFM insulator by a factor of 2$\sim$3 \cite{lichtenstein2001finite}. Upon hole doping, the local moment decreases gradually and drops to zero when hole concentration reaches to 0.2 /f.u.. Meanwhile, the N\'{e}el temperature also decreases as hole concentration increases. This evidence suggests that hole doping can suppress AFM coupling in $\mathrm{NdNiO_2}$. By analyzing the orbital resolved DMFT spectral function at different hole concentrations and temperatures, we find that insulator-metal transition also exists following the AFM-PM transition. As shown in Fig. 8(a), a gap is opened at the Fermi level, representing the undoped $\mathrm{NdNiO_2}$ is an AFM insulator at 300 K when self-doping effect is excluded. The spectral function also indicates that $d_{x^2-y^2}$ orbital is nearly half filled, revealing its nature of doped single-band Hubbard model \cite{kitatani2020nickelate}. When the temperature reaches 1500 K, the $d_{x^2-y^2}$ orbital has a nonzero contribution to the DOS near the fermi energy, which is a sign of bad insulator. Fig. 8(c) shows the spectral function at 300 K with 0.05 /f.u. hole doping, we see that the band gap is closed and a finite peak appears at fermi energy, indicating the system is metallic but its conductivity is limited, i.e., a bad metal phase is obtained which is defined through the shape of electronic spectra A($\omega$) \cite{kotliar2004strongly}. As the hole concentration or temperature continues to increase, $\mathrm{NdNiO_2}$ finally becomes a metal.

\section{\label{sec:level8}Summary}	
In conclusion, we performed DFT+$U$, DMFT calculations, and model analysis to investigate the magnetic properties of $\mathrm{NdNiO_2}$ and $\mathrm{CaCuO_2}$. We found that the magnetic properties of nickelates are significantly affected by both $\Delta_{dp}$ and doping. The large $\Delta_{dp}$ in $\mathrm{NdNiO_2}$ results in a large local moment and spin splitting. It also decreases the AFM coupling. In addtion, the self-doped holes frustrates the AFM. These effects together induce competing magnetic phase in nickelate, which differs from the dominating AFM phase in cuprate. Moreover, the mean-field calculations found that the ground state of 0.2 hole/f.u. doped $\mathrm{NdNiO_2}$, which is in the superconduting state in experiment, is strikingly in a intralayer FM state. This may point to a phase with strong FM fluctuations in the highly doped nickelates, which may be explored in future experiments. The magnetic exchange interaction $J$ estimated by ED of small cluster is 62$\pm$9 meV, comparable to (albeit quantitatively smaller than) that of the cuprates. This magnetic coupling should be quite important for understanding the physical mechanism of supercondutivity in nickelates. In addition, the small magnetic moment observed in neutron scattering experiment should be understood from a dynamical perspective. Other ``faster" methods such as core-level spectroscopy may reveal much larger local moment.

\section{\label{sec:level9}Ackonwledgements}	
We acknowledge financial support from the National Key R$\&$D Program of China (Grant No. 2021YFA0718900, and No. 2017YFA0303602), the Key Research Program of Frontier Sciences of CAS (Grant No. ZDBS-LY-SLH008), the National Nature Science Foundation of China (Grants No. 11974365, No. 12004400, No. 11904373, and No. 51931011), K.C. Wong Education Foundation (GJTD-2020-11), and Ningbo Natural Science Foundation (202003N4363). Calculations were performed at the Supercomputing Center of Ningbo Institute of Materials Technology and Engineering and Vienna Scientific Clusters (VSC).

\appendix

\section{Strain Effect}

	\begin{figure}[h]
	\centering
	\includegraphics[width=0.45\textwidth]{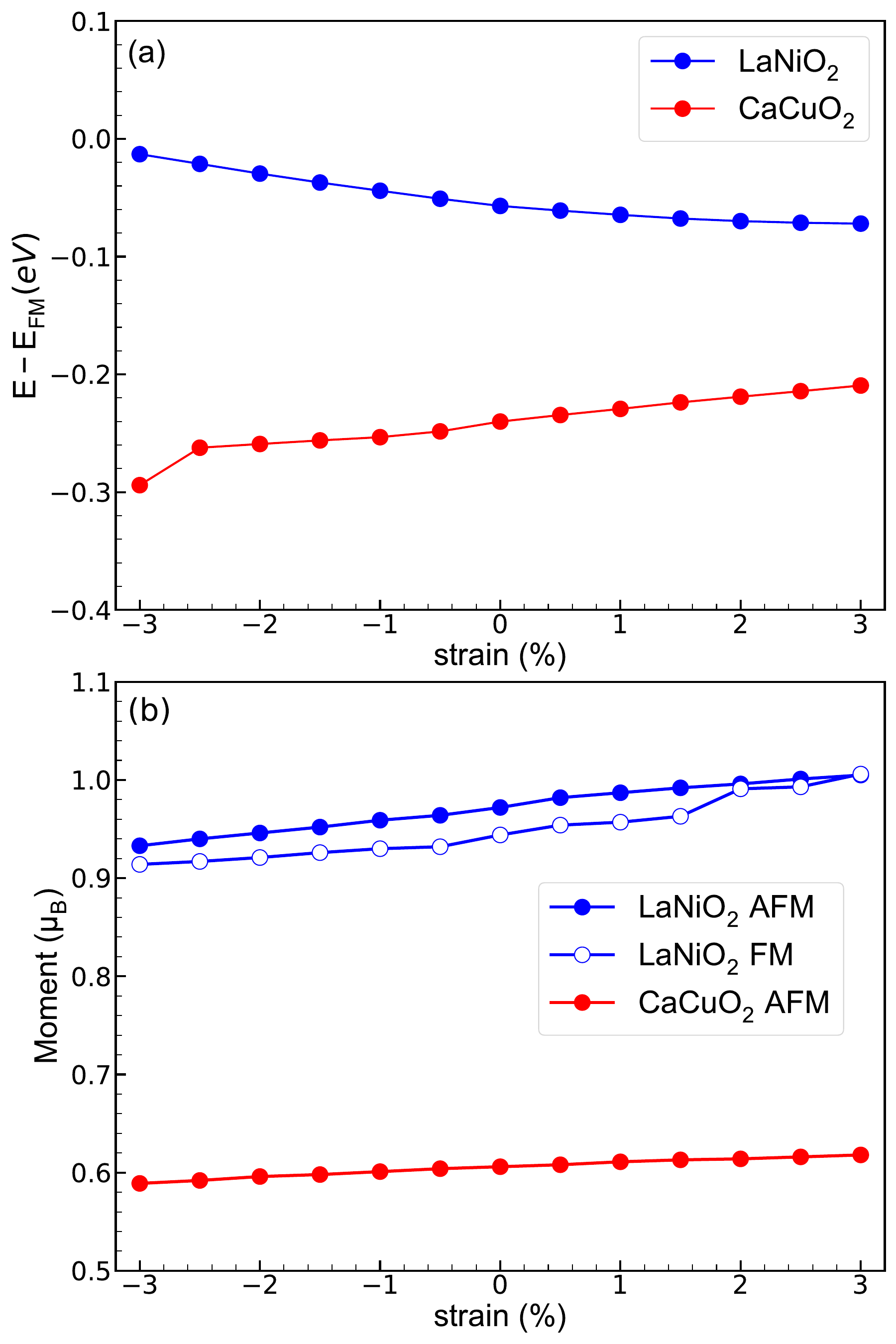}
	\caption{\label{fig:6} HSE calculations of $\mathrm{LaNiO_2}$ and $\mathrm{CaCuO_2}$ at different strains. (a) The energy difference between G-type AFM and FM. (b) The local moment of Ni and Cu for G-type AFM and FM.}
	\end{figure}

To investigate the effect of substrate, we calculate the total energy and the local magnetic moment of $\mathrm{LaNiO_2}$ and $\mathrm{CaCuO_2}$ with strain range for $-3\%$ to 3$\%$. As Fig. 6 shows, the magnetic properties are not significantly influenced by strain, which suggests the effect of substrate is negligible.

\section{Magnetic Order}

	\begin{table}[H]
	\centering
    \setlength{\abovecaptionskip}{0.cm}
    \vspace{-0.4cm}
	\caption{\label{tab:tableS1}Energy comparison for Ni and Nd spin configurations with different $U$.}
	\begin{ruledtabular}
		\begin{tabular}{ccccc}
			&\multicolumn{2}{c}{U-Nd-$f$ = 2.0 eV}&\multicolumn{2}{c}{U-Ni-$d$ = 5.0 eV}\\ \hline
			&Nd FM&Nd G-AFM&Nd C-AFM&Nd A-AFM\\ \hline
			Ni FM&$-$0.059 eV& 0.000 eV&$-$0.010 eV&$-$0.004 eV\\
			Ni G-AFM&$-$0.053 eV&$-$0.046 eV&$-$0.052 eV&$-$0.049 eV\\
			Ni C-AFM&$-$0.337 eV&$-$0.333 eV&$-$0.336 eV&$-$0.334 eV\\
			Ni A-AFM&$-$0.062 eV&$-$0.052 eV&$-$0.059 eV&$-$0.056 eV
		\end{tabular}
		~\\
		~\\
		\begin{tabular}{ccccc}
			&\multicolumn{2}{c}{U-Nd-$f$ = 2.0 eV}&\multicolumn{2}{c}{U-Ni-$d$ = 3.0 eV}\\ \hline
			&Nd FM&Nd G-AFM&Nd C-AFM&Nd A-AFM\\ \hline
			Ni FM&$-$0.348 eV&$-$0.387 eV&$-$0.397 eV&$-$0.391 eV\\
			Ni G-AFM&$-$0.461 eV&0.000 eV&$-$0.359 eV&$-$0.393 eV\\
			Ni C-AFM&$-$0.738 eV&$-$0.733 eV&$-$0.739 eV&$-$0.315 eV\\
			Ni A-AFM&$-$0.445 eV&$-$0.435 eV&$-$0.443 eV&$-$0.332 eV
		\end{tabular}
		~\\
		~\\
		\begin{tabular}{ccccc}
			&\multicolumn{2}{c}{U-Nd-$f$ = 7.0 eV}&\multicolumn{2}{c}{U-Ni-$d$ = 4.0 eV}\\ \hline
			&Nd FM&Nd G-AFM&Nd C-AFM&Nd A-AFM\\ \hline
			Ni FM&$-$0.297 eV&$-$0.147 eV&$-$0.400 eV&$-$0.149 eV\\
			Ni G-AFM&$-$0.309 eV&$-$0.015 eV&$-$0.310 eV&$-$0.016 eV\\
			Ni C-AFM&$-$0.344 eV&$-$0.355 eV&$-$0.290 eV&$-$0.356 eV\\
			Ni A-AFM&$-$0.005 eV&$-$0.004 eV&$-$0.004 eV&0.000 eV
		\end{tabular}
	\end{ruledtabular}
	\end{table}

In the main text, we treat Nd-4$f$ orbitals as core states and the magnetic moment of Nd is ignored for simplicity. Actually, the Nd-4$f$ orbitals contribute large moment and play a important role for further understanding the magnetic behavior of nickelates. Here we calculate the magnetic order by including both of the magnetic moment of Ni and Nd, and the U dependent is also considered. All calculations indicate that the magnetic energy is mainly determined by the magnetic order of Ni.

\section{HSE Calculation}

In Fig. 2, we calculate magnetic moment and total energy at different doping concentrations with DFT+$U$ method. To verify the reliability of our conclusion, we have done the calculation with HSE hybrid functional as shown in Fig. 7, which is consistent with DFT+$U$ results. Therefore, our conclusion is quite robust and will not be changed under a different functional.

	\begin{figure}[htbp]
	\centering
	\includegraphics[width=0.45\textwidth]{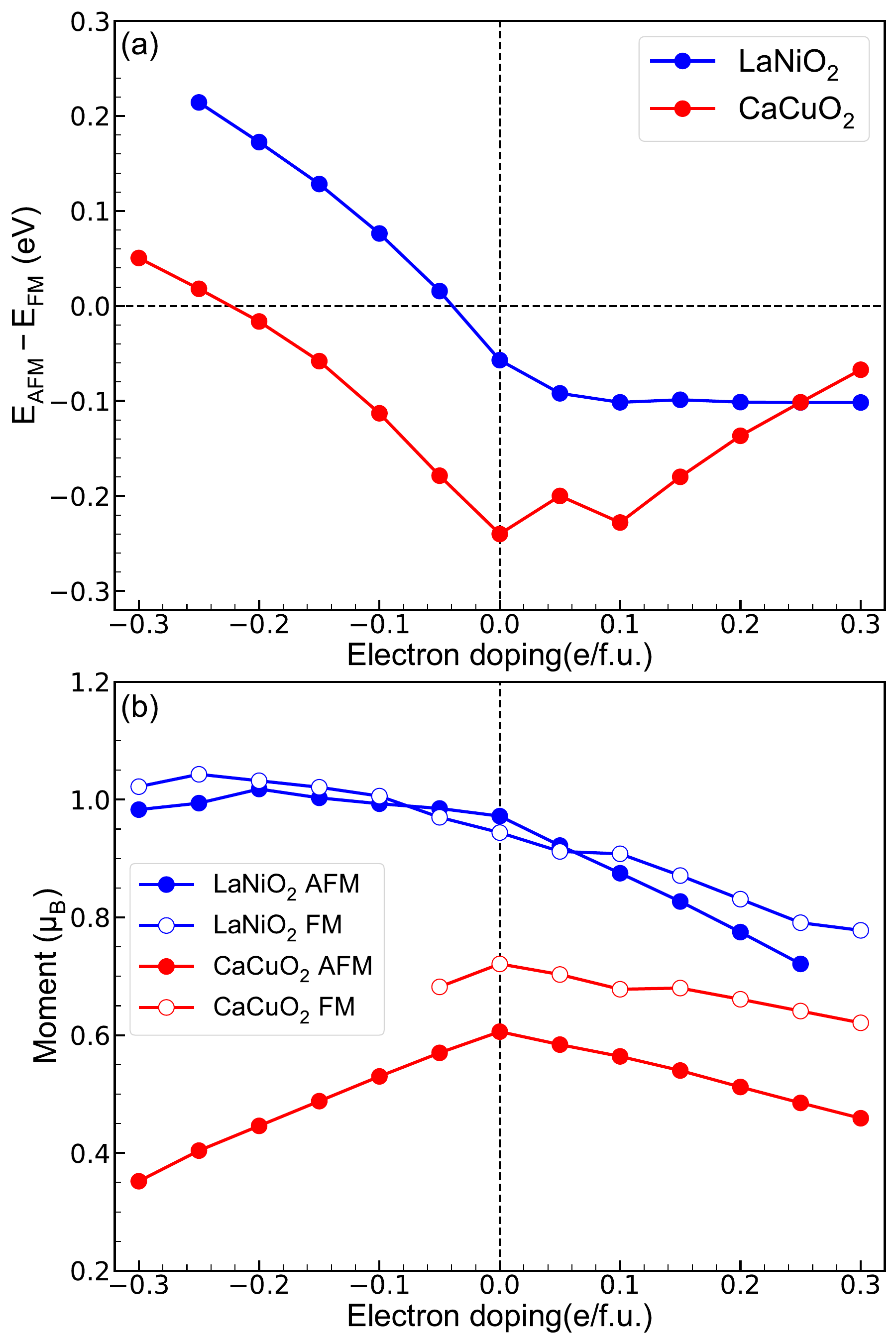}
	\caption{\label{fig:7} HSE calculations of $\mathrm{LaNiO_2}$ and $\mathrm{CaCuO_2}$ at different doping concentrations. (a) The energy difference between G-type AFM and FM. (b) The local moment of Ni and Cu for G-type AFM and FM.}
	\end{figure}

\section{DMFT Calculation}

	\begin{figure*}[htbp]
	\centering
    \vspace{-0.2cm}
	\includegraphics[width=0.7\textwidth]{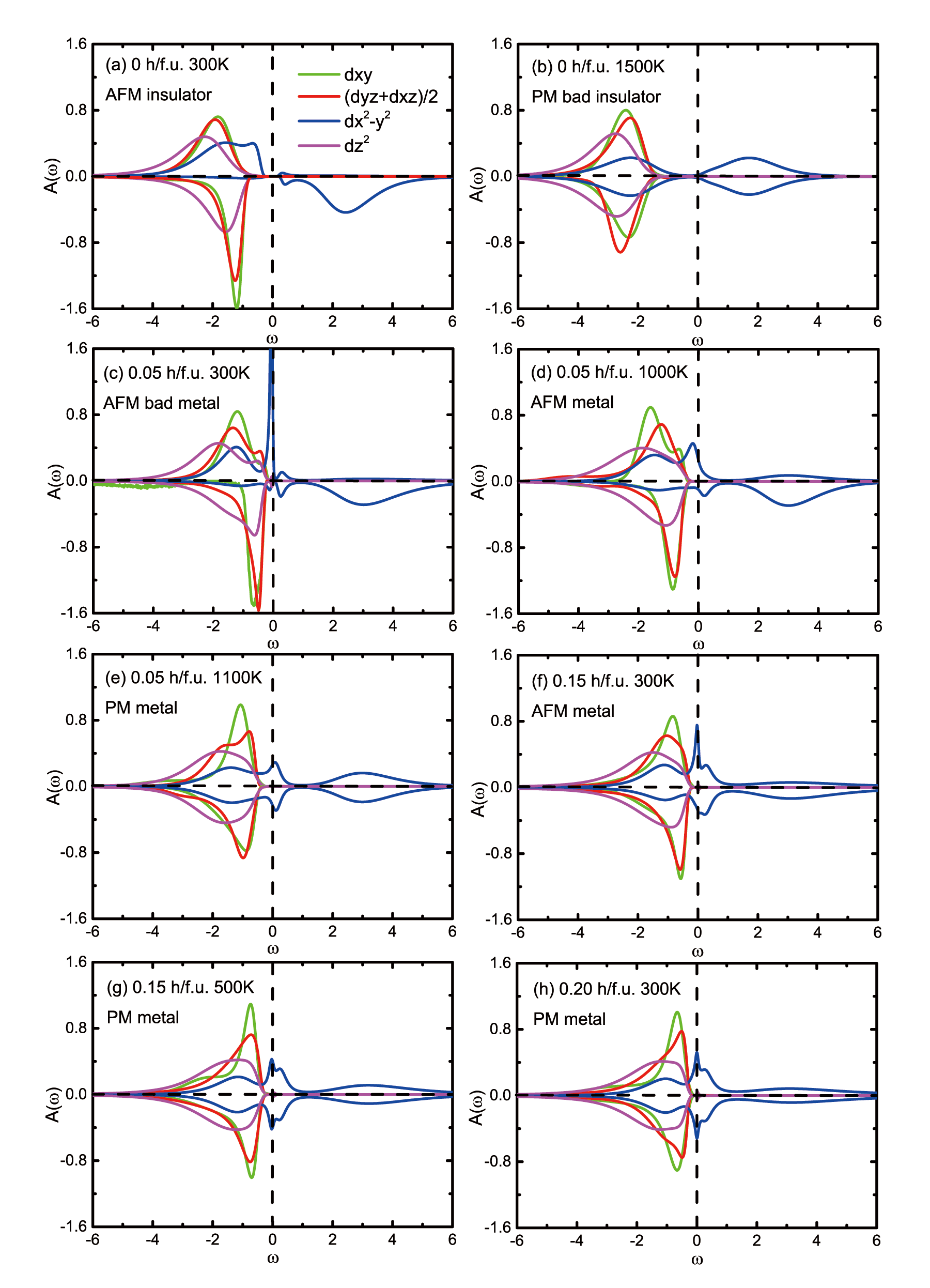}
	\caption{\label{fig:9} Orbitally resolved spectral functions calculated by DFT+DMFT at different temperature and doping concentration. The positive and negative indicate spin-up and -down channels.}
	\end{figure*}

	\begin{figure}[htbp]
	\centering
	\includegraphics[width=0.4\textwidth]{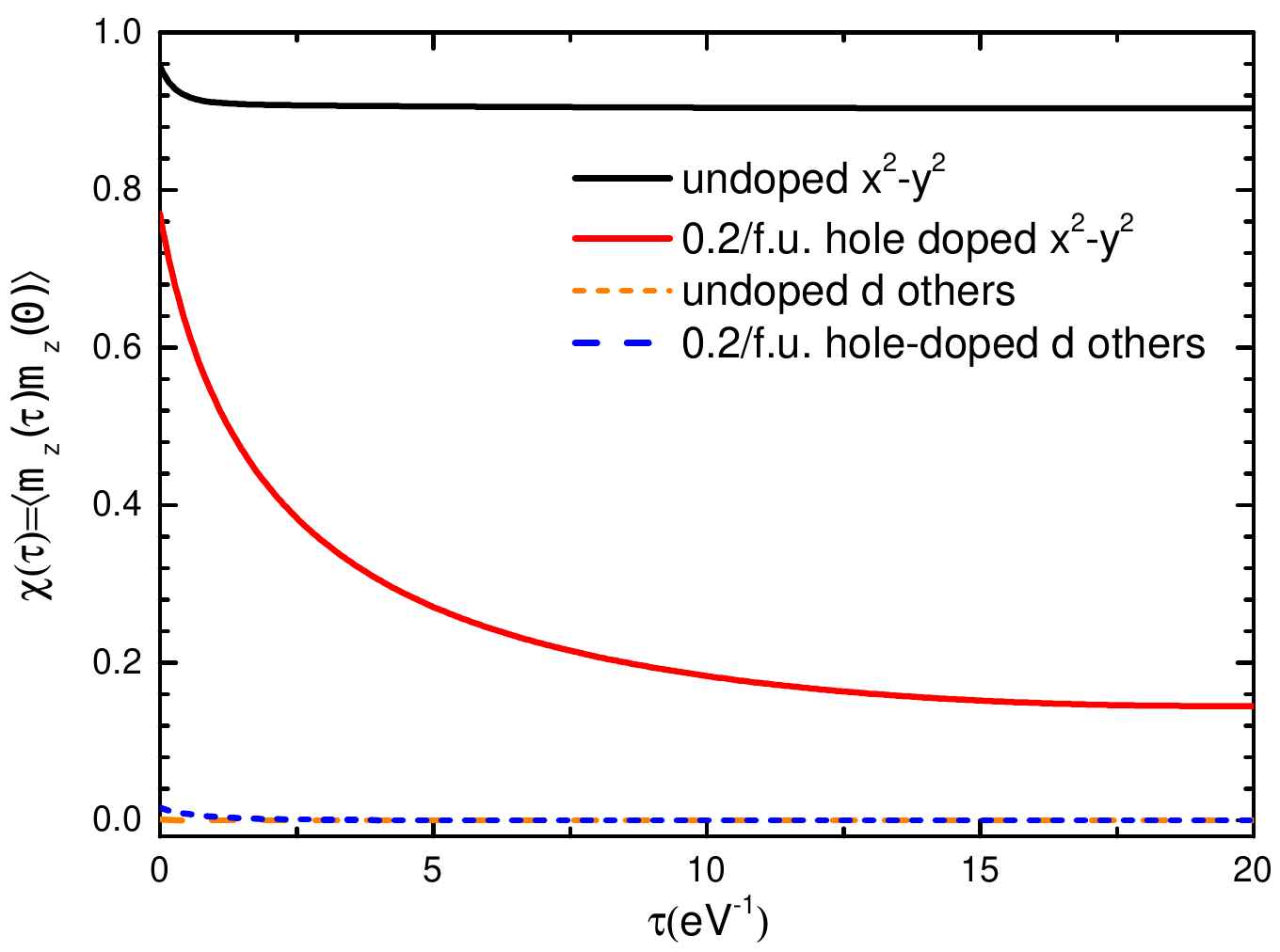}
	\caption{\label{fig:8} Orbitally resolved local spin correlation functions $\chi(\tau)=\left\langle\hat{m_z}(\tau)\hat{m_z}(0)\right\rangle$ calculated by DFT+DMFT at $T$ = 290 K in the density-density form in $d$-only model, $\ \beta/2\approx $ 20\ eV$^{-1}$.}
	\end{figure}

To show more details about the phase diagram of $\mathrm{NdNiO_2}$ calculated by DMFT (see Fig. 5), especially for the definition of the conductivity (bad metal phase) and also the magnetic moment, we plot the spectral functions at different temperature and hole doping concentration in Fig. 8. As the orbital resolved local spin correlation functions (see Fig. 4) shows in Sec. III D, the dynamical screening effect emerged in Ni-$d$ orbital. To compare with the Kanamori form in the main text, we also calculate the spin correlation functions in density-density form as shown in Fig. 9, which is consist with the conclusions above.

\clearpage

\bibliography{document}

\end{document}